# Calculation of the orthorhombic *E*-parameter in EPR for $d^3$ spin systems

Th. W. Kool

Van 't Hoff Institute for Molecular Sciences, University of Amsterdam

May 2011

**Abstract**

Third order perturbation theory is used to calculate the orthorhombic spin-Hamiltonian zero-field *E*-term for octahedrally surrounded $d^3$ spin systems ($S = 3/2$) in strong axial fields in the presence of a weak orthorhombic component.

**Introduction**

In this paper we present results of analytical third-order perturbation calculations of the zero-field splitting term *E* for octahedrally surrounded $d^3$ ($S = 3/2$) systems, in the presence of strong and moderate axial fields and a weak orthorhombic perturbation, i.e. for $|D| \geq h\nu$ and $|E| \leq h\nu$, with ν the frequency of typically an EPR experiment. Expressions for $d^5$ ($S = 5/2$) systems were already derived before i.e., for the $Fe^{3+}$-$V_O$ system in $SrTiO_3$ [1, 2].

The ground state of $d^3$ ($S = 3/2$) ions in an octahedral field is $^4A_2$ [3-6]. All excited states are lying higher in energy by amounts large compared with the spin-orbit coupling. The $^4A_2$ state is connected through spin-orbit coupling with the excited $T_{2g}$ states only [2-7]. Use of second-order perturbation theory gives a *g*-value slightly less than 2. For axially distorted (tetragonal or trigonal) octahedrally surrounded $d^3$ spin systems the following spin-Hamiltonian is used [1-7]:

$$\mathcal{H} = S \cdot \bar{D} \cdot S + \mu_B H \cdot \bar{g} \cdot S \qquad (1)$$

The first term represents the zero-field splitting and the second term the Zeeman interaction. The spin degeneracy of the $^4A_2$ state is partly removed into two Kramers' doublets separated by $|2D|$. The principal contribution to the *g*-shifts is caused by mixing with the excited $^4T_2$ state, which is split into an orbital singlet and an orbital doublet state. Mixing with the $^4T_2$ state, interaction with other levels and spin-spin coupling leads to the observed zero-field splitting. If the zero-field splitting $|2D|$ is much larger than the Zeeman term, only one EPR transition within the Kramers doublet with $M_S = |\pm 1/2\rangle$ levels is observed. The angular dependence of the effective *g*-values can be obtained by perturbation theory within the $^4A_2$ term using as basis the $M_s = |\pm 3/2\rangle$ and $M_s = |\pm 1/2\rangle$ wave functions and is given by:

$$\mathcal{H} = D[S_z^2 - \tfrac{1}{3}S(S+1)] + g_\parallel \mu_B H S_z \cos\alpha + g_\perp \mu_B H S_x \sin\alpha, \qquad (2)$$

with α the angle between the magnetic field $\vec{H}$ and the centre axis $z$.

This is justified because the excited $T_2$ states are lying much higher in energy. The effective g-values in first order are given by $g_\parallel^{eff} = g_\parallel$ and $g_\perp^{eff} = 2g_\perp$ with an effective spin of $S' = \frac{1}{2}$ [1-7]. The angular dependence of the effective g-values are altered in second- and third-order perturbation theory by [1-2, 8-9]:

$$g^{eff}(\alpha) = \left(g_\parallel^2 \cos^2\alpha + 4g_\perp^2 \sin^2\alpha\right)^{\frac{1}{2}} \left[1 - \frac{3}{4}\left[\frac{g_\perp \mu_B H}{2D}\right]^2 F(\alpha)\right], \text{ where} \qquad (3)$$

$$F(\alpha) = \sin^2\alpha \left[\frac{(4g_\perp^2 + 2g_\parallel^2)\sin^2\alpha - 2g_\parallel^2}{(4g_\perp^2 - g_\parallel^2)\sin^2\alpha + g_\parallel^2}\right]. \qquad (4)$$

α is the angle between the magnetic field $H$ and the axial centre axis $z$ of the $d^3$ system and $\mu_B$ is the Bohr magneton.

Hence $g^{eff}(0°) = g_\parallel$

**Spin Hamiltonian for an orthorhombic $S = 3/2$ system**

The spin-Hamiltonian for octahedrally surrounded orthorhombic distorted $d^3$ systems ($S = 3/2$) need an extension with the following term [1, 2]:

$$E(S_x^2 - S_y^2). \qquad (5)$$

Using θ and δ as the polar angles of the magnetic field $\vec{H}$ with respect to the main axis z of the local centre we then obtain the following spin-Hamiltonian expressed in spin operators $S_z$ and $S_\pm = S_x \pm iS_y$,

$$\mathcal{H} = D\left(S_z^2 - \frac{1}{4}\right) + \frac{1}{2}E(S_+^2 + S_-^2)$$
$$+ \mu_B H\left[g_z S_z \cos\theta + \frac{1}{2}g_x(S_+ + S_-)\sin\theta\cos\delta - \frac{1}{2}ig_y(S_+ + S_-)\sin\theta\cos\delta\right]$$
$$(6)$$

where it is understood that the level of zero energy is chosen so that the first term of eq. (6) is zero for the $S_z = \pm 1/2$ doublet. For systems with $|D| \gg h\nu$ the first term is taken as zero order Hamiltonian $\mathcal{H}_0$ and the other crystal field terms and the Zeeman splitting are treated as a perturbation $\mathcal{H}_1$. For values $h\nu/|2D| \geq 0.25$ one has to proceed with exact numerical computer calculations [3]. Using as basis the $|\pm 3/2\rangle$ and $|\pm 1/2\rangle$ wave functions, the matrix of the Hamiltonian is given by:

| ℋ | $|+3/2\rangle$ | $|+1/2\rangle$ | $|-1/2\rangle$ | $|-3/2\rangle$ |
|---|---|---|---|---|
| $\langle+3/2|$ | 2D+3A | $\sqrt{3}\cdot B^*$ | $\sqrt{3}\cdot E$ | 0 |
| $\langle+1/2|$ | $\sqrt{3}\cdot B$ | A | $2B^*$ | $\sqrt{3}\cdot E$ |
| $\langle-1/2|$ | $\sqrt{3}\cdot E$ | 2B | -A | $\sqrt{3}\cdot B^*$ |
| $\langle-3/2|$ | 0 | $\sqrt{3}\cdot E$ | $\sqrt{3}\cdot B$ | 2D-3A |

with:  $A = \tfrac{1}{2}g_z\mu_B H\cos\theta$
$B = \tfrac{1}{2}\mu_B H\sin\theta(g_x\cos\delta + ig_y\sin\delta)$
$B^* = \tfrac{1}{2}\mu_B H\sin\theta(g_x\cos\delta - ig_y\sin\delta)$ and
$\mu_B$ the Bohr magneton.

The eigenvalues of $\mathcal{H}_0$ are doubly degenerate and $\mathcal{H}_1$ couples via B and $B^*$ between eigenstates within the same eigenspace with energy zero. The basis of this eigenspace has first to be adapted such that the submatrix of ℋ within the eigenspace $|\pm 1/2\rangle$ becomes diagonal.

The new wave functions then become:

$|\gamma\rangle = c_1|+\tfrac{1}{2}\rangle + c_2|-\tfrac{1}{2}\rangle$ and $|\delta\rangle = -c_2|+\tfrac{1}{2}\rangle + c_1^*|-\tfrac{1}{2}\rangle$, where  (7)

$c_1 = \left[\dfrac{B^*}{B}\right]^{1/2} \cdot \left[\dfrac{(A^2+|2B|^2)^{1/2}+A}{2(A^2+|2B|^2)^{1/2}}\right] = \left[\dfrac{P+A}{2P}\right] \cdot \left[\dfrac{B^*}{B}\right]^{1/2}$ and

$c_2 = \left[\dfrac{P-A}{2P}\right]^{1/2}$

The energies are: $W_{\gamma,\delta} = \pm(A^2 + |2B|^2)^{1/2} = \pm P$.  (8)

The above matrix transforms within the new basis as,

| ℋ | $|+3/2\rangle$ | $|\gamma\rangle$ | $|\delta\rangle$ | $|-3/2\rangle$ |
|---|---|---|---|---|
| $\langle+3/2|$ | 2D+3A | $c_1\sqrt{3}\cdot B^* + c_2\sqrt{3}\cdot E$ | $c_1^*\sqrt{3}\cdot E - c_2\sqrt{3}\cdot B^*$ | 0 |
| $\langle\gamma|$ | $c_1^*\sqrt{3}\cdot B + c_2\sqrt{3}\cdot E$ | +P | 0 | $c_1^*\sqrt{3}\cdot E + c_2\sqrt{3}\cdot B^*$ |
| $\langle\delta|$ | $c_1\sqrt{3}\cdot E - c_2\sqrt{3}\cdot B$ | 0 | -P | $c_1\sqrt{3}\cdot B^* - c_2\sqrt{3}\cdot E$ |
| $\langle-3/2|$ | 0 | $c_1\sqrt{3}\cdot E + c_2\sqrt{3}\cdot B$ | $c_1^*\sqrt{3}\cdot B - c_2\sqrt{3}\cdot E$ | 2D-3A |

If $E = 0$, second-order perturbation does not split the $|\gamma\rangle$ and $|\delta\rangle$ levels further. With $E \neq 0$, an additional splitting is caused by the admixture of $|\pm 3/2\rangle$ levels into the ground states $|\gamma\rangle$ and $|\delta\rangle$ and is given by:

$$\Delta W_\gamma^{(2)} = -\sum_{\gamma \neq 3/2} \frac{\langle +3/2|\mathcal{H}|\gamma\rangle\langle\gamma|\mathcal{H}|+3/2\rangle}{(W_{+3/2}-W_\gamma)} = -\sum_{\gamma \neq -3/2} \frac{\langle -3/2|\mathcal{H}|\gamma\rangle\langle\gamma|\mathcal{H}|-3/2\rangle}{(W_{-3/2}-W_\gamma)} \quad (10)$$

This gives:

$$\Delta W_\gamma^{(2)} = \frac{-6\mu_B E(c_1^* B + c_1 B^*)}{2D} = -\Delta W_\delta^{(2)}$$

$$\Delta W_\gamma^{(2)} - \Delta W_\delta^{(2)} = \frac{-6\mu_B E(c_1^* B + c_1 B^*)}{D} = \frac{-6E}{D} \cdot \frac{\sin^2\theta(g_x^2\cos^2\delta - g_y^2\sin^2\delta)}{(g_z^2\cos^2\delta + 4g_\perp^2\sin^2\delta)^{1/2}} \quad (11)$$

In third-order perturbation theory $\Delta W_\gamma^{(3)}$ is given by [10],

$$\Delta W_\gamma^{(3)} = \sum_{m \neq \gamma,\delta} \frac{|H_{\gamma m}|^2}{(W_m - W_\gamma)^2}(\mathcal{H}_{mm} - \mathcal{H}_{\gamma\gamma}) + \sum_{m \neq \gamma,\delta} \sum_{p \neq m,\gamma,\delta} \frac{\mathcal{H}_{mp}\mathcal{H}_{p\gamma}}{(W_\gamma - W_p)(W_\gamma - W_m)} \quad (12)$$

where $m$ is the ground state.
The second term is equal to zero, because according to $\sum_{m \neq \gamma,\delta}$ the wave functions $\gamma$ and $\delta$ may not be used as variables and matrix elements $\mathcal{H}_{mp} \cdot \mathcal{H}_{p\gamma}$ give zero.
The first term gives the following result,

$$\Delta W_\gamma^{(3)} = \frac{3BB^*}{2D} - \frac{3PBB^*}{4D^2} + \frac{9A^2 BB^*}{4D^2 P} \quad (13)$$

For systems with $E \ll D$, the terms in $E/D$, $E/D^2$, $E^2/D^2$ and $E^2/D$ are small and are therefore omitted, therefore $\Delta W_\gamma^{(3)} = -\Delta W_\gamma^{(3)}$.
We are now able to give the expression for the g-value, which is as follows,

$$g^{eff}\mu_B H = W_\gamma + \Delta W_\gamma^{(2)} + \Delta W_\gamma^{(3)} - W_\delta - \Delta W_\delta^{(2)} - \Delta W_\delta^{(3)} = 2P + 2\Delta W_\gamma^{(2)} + 2\Delta W_\gamma^{(3)} \quad (15)$$

After some calculations we obtain:

$$g^{eff}(\theta) = (g_z^2\cos^2\theta + 4g_\perp^2\sin^2\theta)^{\frac{1}{2}}\left[1 - \frac{3}{4}\left[\frac{g_\perp\mu_B H}{2D}\right]^2 F(\theta)\right] - \frac{6E}{D}\sin^2\theta \frac{g_x^2\cos^2\delta - g_y^2\sin^2\delta}{(g_z^2\cos^2\theta - 4g_\perp^2\sin^2\theta)^{1/2}}$$

$$(16)$$

where
$F(\theta) = \sin^2\theta \left[\frac{(4g_\perp^2 + 2g_z^2)\sin^2\theta - 2g_z^2}{(4g_\perp^2 - g_z^2)\sin^2\theta + g_z^2}\right]$ and $g_\perp^2 = g_x^2\cos^2\delta + g_y^2\sin^2\delta$

The expression for the axial field (with $E = 0$) is given by eq. (3).

The results obtained above were used in a study of the non-cubic $Fe^{5+}$ centre in $SrTiO_3$ [2, 11]. In the extreme case, where the Zeeman term dominates the axial field, see eq. (1), the solutions are given by references [4, 7]. Examples of the latter are the $SrTiO_3$:$Mo^{3+}$ [12] and $SrTiO_3$:$Cr^{3+}$ systems [2, 13].